\documentclass[aps,prstab,floatfix,reprint,groupedaddress,showpacs,showkeys,longbibliography,a4paper]{revtex4-1}

\usepackage{graphicx}
\usepackage{subfigure}

\begin{document}


\title{Effects of rf breakdown on the beam in the Compact Linear Collider
prototype accelerator structure}

\author{A. Palaia}
\email[Corresponding author: ]{andrea.palaia@physics.uu.se}
\author{M. Jacewicz}
\author{R. Ruber}
\author{V. Ziemann}
\affiliation{Department of Physics and Astronomy, Uppsala
University, Uppsala, Sweden}

\author{W. Farabolini}
\affiliation{CEA IRFU Centre d'Etudes de Saclay, France}


\begin{abstract}
Understanding the effects of RF breakdown in high-gradient accelerator
structures on the accelerated beam is an extremely relevant aspect in the
development of the Compact Linear Collider (CLIC) and is one of the main issues
addressed at the Two-beam Test Stand at the CLIC Test Facility 3 at CERN. During
a RF breakdown high currents are generated causing parasitic magnetic fields
which interact with the accelerated beam affecting its orbit. The beam energy is
also affected because the power is partly reflected and partly absorbed thus
reducing the available energy to accelerate the beam.
We discuss here measurements of such effects observed on an electron beam
accelerated in a CLIC prototype structure. Measurements of the trajectory of
bunch-trains on a nanosecond time-scale showed fast changes in correspondence of
breakdown which we compare with measurements of the relative beam spots on a
scintillating screen. We identify different breakdown scenarios for which we
offer an explanation based also on measurements of the power at the input and
output ports of the accelerator structure. Finally we present the distribution
of the magnitude of the observed changes in the beam position and we discuss its
correlation with RF power and breakdown location in the accelerator structure.
\end{abstract}


\maketitle

\section{\label{sec:intro}Introduction}

The Compact Linear Collider (CLIC) study aims to develop a linear accelerator to
accelerate electrons and positrons to TeV scale energies~\cite{clic_cdr_1}. In
order to achieve this goal in a realistic and cost-effective way the loaded
accelerating gradient needs to be as high as 100~MV/m. This high gradient
precludes the use of superconducting accelerator structures~\cite{Furuta:2006jj}
and therefore uses room-temperature X-band (12~GHz) technology.
\par The extremely high accelerating gradient entails the presence of electric
field strengths on the inner surfaces of the structures in excess of
200~MV/m~\cite{TD24:grudiev} which causes field emission and even discharges,
so-called RF breakdown.
The field emission is triggered from nano-metric imperfections on the structure
walls which is observed as dark current during normal operation.
It can randomly initiate an avalanche process which results in a RF breakdown.
Breakdowns are a severe problem for the operational reliability of any
accelerator, because an accelerator structure with a discharge is effectively
``shorted" and the RF power is partly reflected and partly absorbed in the
process, thus reducing the available power to accelerate the beam.
A second effect is that the strong currents flowing during the discharge cause
fields that deflect the beam. These processes were investigated earlier at SLAC
where dark currents and breakdown in both waveguides and accelerator structures
have been simulated~\cite{Dolgashev:823150, Dolgashev:2002vv, Bane:2005kk}.
It was found that transverse momentum can be transferred to the beam and its
magnitude was experimentally measured up to 30~keV/c~\cite{Adolphsen:2005iy,
Dolgashev:2004}.

In this report we present the first study of the effect of discharges on an
accelerated beam in the Two-beam Test Stand~\cite{Ruber:2008} in the CLIC Test
Facility CTF3 at CERN~\cite{Geschonke:559331}. We focus on the measurements of
transverse kicks to the beam due to RF breakdown in a CLIC prototype accelerator
structure, carried on during the CTF3 2012 run.
\par The remainder of this report is organised as follows. First, we describe
the experimental set-up, beam characteristics and diagnostics installed and used
in the experiment. We then discuss the effect of the acceleration in the CLIC
prototype accelerator structure on a single bunch-train. Afterwards we present
evidence for RF breakdown effects on the beam, showing and discussing
measurements of a few selected examples of bunch-trains accelerated in a CLIC
prototype accelerator structure while a breakdown occurs. Finally, the
methodology used for the analysis of the examples of breakdown presented is
applied to a bigger data set for which we discuss the distribution of transverse
kick magnitude and its correlation with RF power and breakdown location in the
accelerator structure.

\section{CLIC and CTF3} 

The experiments we discuss were performed at CTF3 which was constructed in a
collaborative effort in order to address several key issues of the CLIC design:
the discharges in the accelerator structures is one of them and the experimental
verification of the two-beam acceleration scheme is another. The latter is based
on extraction of RF power from a moderate energy, high-current electron beam
called drive beam, and transfer of such power to a low-current beam called main
beam which is accelerated. The drive beam is decelerated in so-called Power
Extraction and Transfer Structures (PETS) providing multi-MW 12~GHz radio
frequency at the expense of the drive beam energy~\cite{PhysRevSTAB.14.081001}.
This setup is mimicked in CTF3 where the drive beam is passed through an
experimental section, the Two-beam Test Stand~\cite{Ruber:2008}, in parallel to
a second low-current beam, so-called probe beam, which is then accelerated in a
prototype accelerator structure using the energy extracted from the drive beam.
\par
\begin{figure}
  \includegraphics[width=\linewidth]{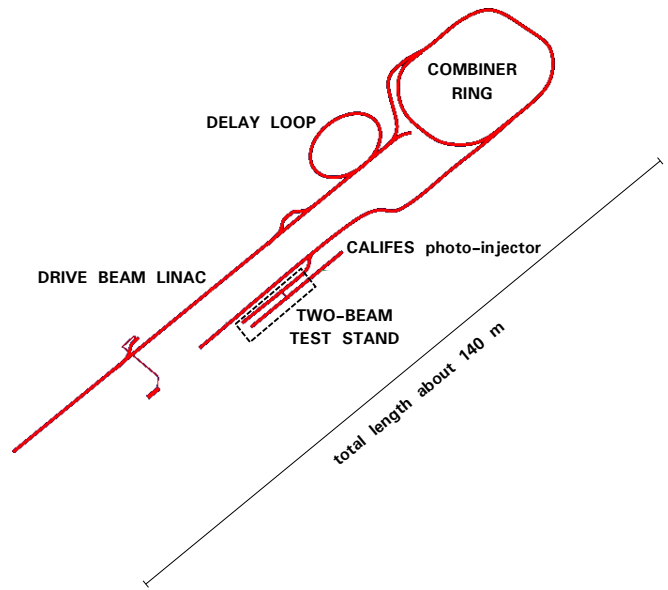}
  \caption{\label{fig:ctf3}Sketch of the CLIC Test Facility 3 at CERN. An
  electron drive beam is accelerated in a normal-conducting linac and its bunch
  frequency and current multiplied by means of a delay loop and a combiner ring.
  The beam is then sent to the Two-beam Test Stand experimental hall where the
  CALIFES linac is also installed to provide a second electron beam
  which is as well sent to the Two-beam Test Stand probe beam line for two-beam acceleration
  experiments.}
\end{figure}

 The Two-beam Test Stand is part of the CTF3 complex sketched in
Fig.~\ref{fig:ctf3} where a 4~A drive beam consisting of a 1.2~$\mu s$ long
train of electron bunches extracted from a thermionic gun at 1.5~GHz is
accelerated in a normal conducting linac to an energy of about 150~MeV. Its
bunch frequency and its current are multiplied up to a maximum of 12~GHz and
28~A, respectively, by means of a delay line called delay loop and a ring called
combiner ring, in order to build the time structure needed to efficiently
produce 12~GHz RF power in the PETS. The probe beam is generated on a
photocathode in a linac called CALIFES~\cite{Farabolini:1363658} and then sent
to the Two-beam Test Stand for two-beam acceleration experiments. A comparison
between CLIC and CTF3 beam parameters is given in
Table~\ref{tab:clic_vs_ctf3_params}.
The final drive beam time structure at CTF3 is the same as the CLIC one although
in CTF3 it is achieved using a different combination of delay lines and rings
than in CLIC. The beam current and energy (before deceleration) are lower in
CTF3 than in CLIC. The CLIC main beam and the CTF3 probe beam have a similar
time structure but different energy (before acceleration) and current, the
latter being limited by beam loading and space charge in the accelerator
structures used in the CALIFES linac~\cite{Mosnier:1078543}.
The two-beam acceleration was successfully demonstrated at CTF3 in
2010~\cite{Ruber2010} and afterwards even beyond the CLIC nominal requirements
of 100~MV/m~\cite{Skowronski:ipac11}. Nevertheless other experimental aspects
are addressed at the Two-beam Test Stand which are connected to the acceleration
technology.

\begin{table}
\caption{\label{tab:clic_vs_ctf3_params}Summary of CLIC and CTF3 beam
parameters.}
\begin{ruledtabular}
\begin{tabular}{lcc}
 & \textbf{CLIC nominal} & \textbf{CTF3 2012} \\
 \textbf{Drive beam} &  &  \\
 ~~current & 101~A & 28~A \\
 ~~energy & 2.4~GeV & 150~MeV\\
 ~~bunch frequency & 12~GHz & 12~GHz \\
 \textbf{Main/Probe beam} &  &  \\
 ~~current & 1~A & 0.2~A \\
 ~~energy & 9~GeV & 180~MeV \\
 ~~bunch frequency & 2~GHz & 1.5~GHz \\
 ~~bunch charge & $>$ 0.6~nC & 0.085 - 0.6~nC \\
 ~~bunches per pulse & 312 & 1 - 225 \\
 ~~repetition rate & 50~Hz & 0.8 - 5~Hz \\
 ~~transverse emittance & $<$ 0.6 mm mrad & $8\pi$ mm mrad \\
\end{tabular}
\end{ruledtabular}
\end{table}

\begin{table}
\caption{\label{tab:td24_params}Main parameters of the CLIC prototype
accelerator structure installed in the Two-beam Test Stand.}
\begin{ruledtabular}
\begin{tabular}{lc}
 name & TD24\_vg1.8\_disk \\
 frequency & 11.995 GHz \\
 number of cells & 24 + input cell + output cell \\
 length & 22.77 cm \\
 filling time & 64.55 ns \\
 inner radius & 3.15/2.35 mm \\
 group velocity & 1.617/0.811 \% of c \\
 phase advance/cell & 2/3 $\pi$ \\
 input power (unloaded) & 46.55 MW for 100 MV/m \\
 input power (loaded) & ~64 MW for 100 MV/m \\
\end{tabular}
\end{ruledtabular}
\end{table}

\section{\label{sec:setup}Experimental set-up}

\begin{figure*}
  \includegraphics[width=\linewidth]{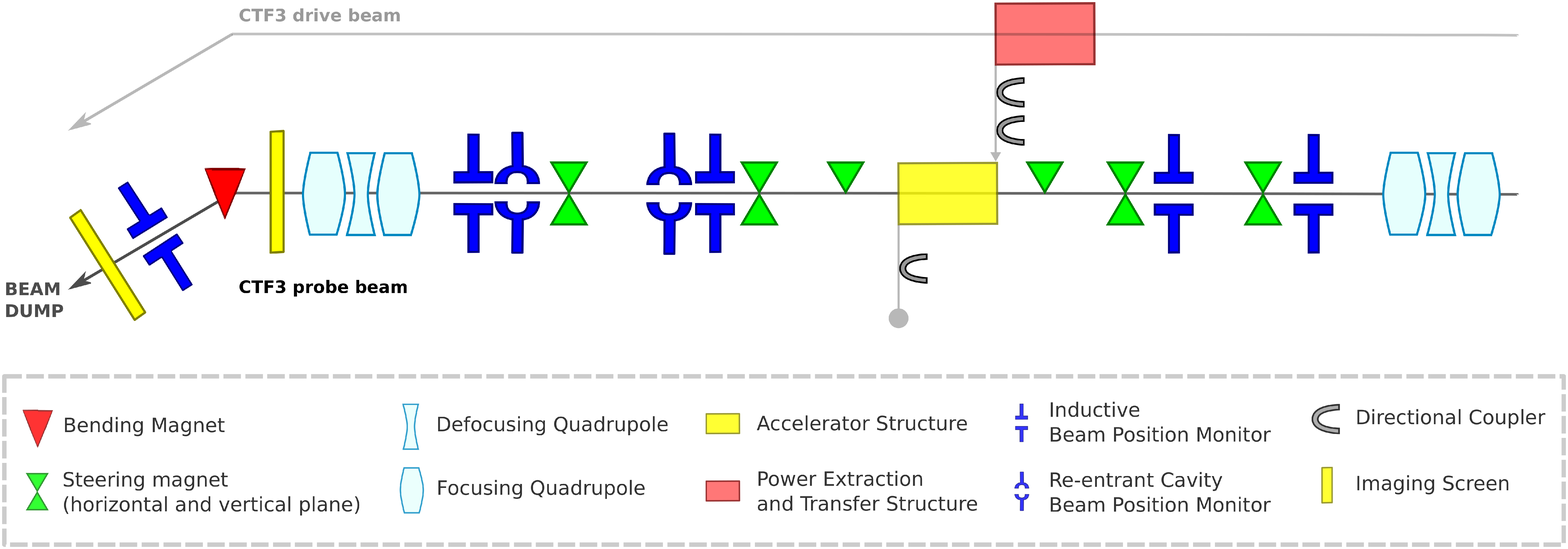}
  \caption{\label{fig:probebeam}Sketch of the probe beam line of the Two-beam
  Test Stand at the CLIC Test Facility 3. The CLIC prototype accelerator
  structure is fed with RF power produced in the Power Extraction and Transfer
  Structure installed in the drive beam line. The probe beam line is equipped
  with two quadrupole triplets, steering magnets, inductive and cavity Beam
  Position Monitors, removable imaging screens and directional couplers at the
  input and output ports of the accelerator structure. A spectrometer line is
  used to measure the beam energy just before the beam dump.}
\end{figure*}

The Two-beam Test Stand consists of two parallel beam lines designed to test
power extraction from the CTF3 drive beam in a CLIC prototype Power Extraction
and Transfer Structure and its transfer through a waveguide network to a probe
beam which is accelerated in a CLIC prototype accelerator structure. Both beam
lines consist of a 11~m long straight section ending with a 1.6~m long
spectrometer line where the beam energy is measured before the beam dump. Both
beam lines are equipped with five inductive beam position
monitors~\cite{Gasior2003}, two upstream and three downstream of each structure,
the last one being in the spectrometer line. The probe beam line sketched in
Fig.~\ref{fig:probebeam} is also equipped with two cavity beam position
monitors~\cite{Simon2009} downstream of the accelerator structure. They were
recently installed to resolve fast changes in the beam position thanks to their
bandwidth of 600~MHz centred at a frequency of 6~GHz which makes them
insensitive to the low frequency noise visible on the inductive beam position
monitor signals.

Both beam lines are equipped with removable imaging screens which are used to
measure the beam spot either just before or after the dipole magnet which bends
the beam in the spectrometer line. The screen in the spectrometer line of the
probe beam is a high sensitivity fluorescent screen whereas at the end of the
straight section of the beam line, before the spectrometer, both an optical
transition radiation and a scintillating screen~\cite{Farabolini:2008} are
available. The measurements discussed in this paper are based on the latter.

\begin{figure*}
\includegraphics[width=\linewidth]{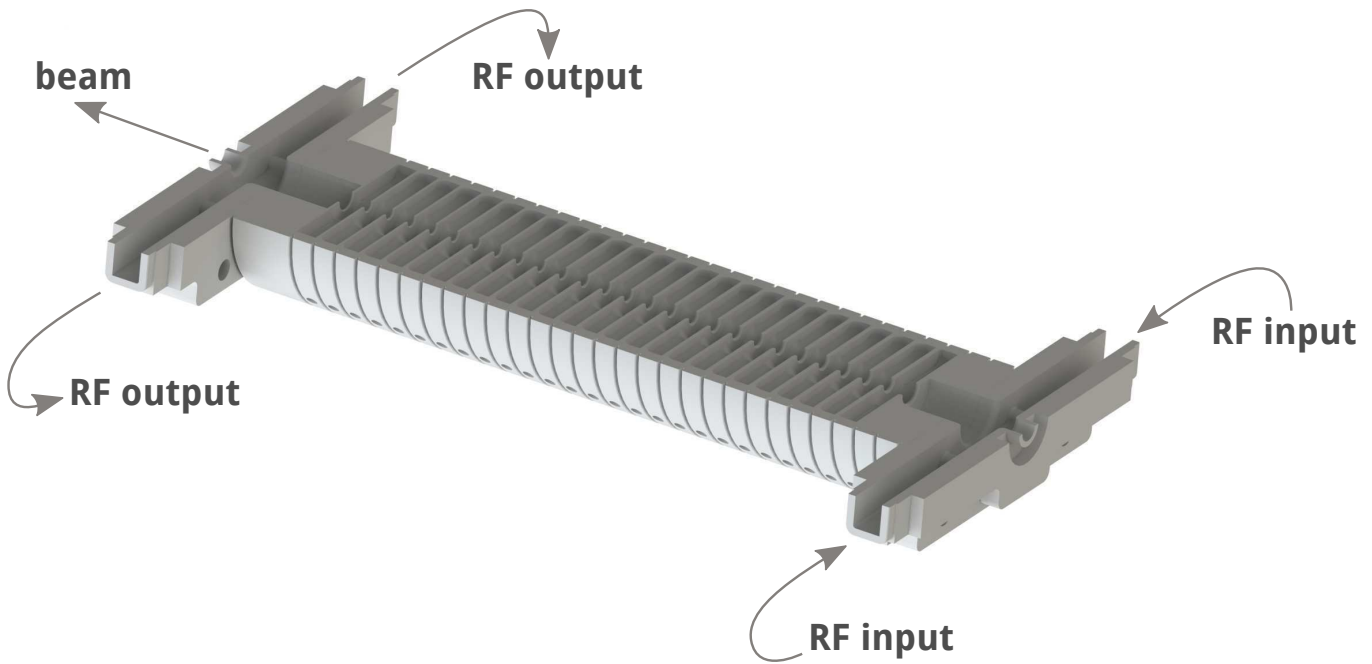}
 \caption{\label{fig:acs_cutout}Longitudinal cross section of a 3D model of the
 CLIC prototype accelerating structure installed in the Two-beam Test Stand in
 2012.}
\end{figure*}

The accelerator structure installed in the beam line is a CLIC prototype. It is
a X-band travelling wave resonant structure made of copper. It consists of 24
regular cells, i.e.\ cells providing nominal field gradient, plus 1 input and 1
output matching cells. The phase advance per cell is $2/3\ \pi$. Its total
length is 22.77~cm. It has a small aperture and a strong linear tapering - group
velocity varies from 1.62\% in the first cell to 0.81\% of the speed of light in
the last cell~\cite{TD24:grudiev}. More specifications are given in
Table~\ref{tab:td24_params} and a cut-out of a 3D model of the structure is
shown in Fig.~\ref{fig:acs_cutout}. No high-order modes damping material are
installed in this prototype. It is equipped with directional couplers at its
input and output ports. Their output signals are sent to diode detectors and IQ
demodulators and then calibrated for power and phase measurements.

Every time that a drive beam bunch-train is sent into the PETS it excites
electromagnetic radiation peaked at 12~GHz, which is sent to the accelerator
structure in the probe beam through a waveguide network. Such a RF pulse is used
to accelerate a probe beam bunch-train that is sent into the structure
synchronous with the RF. The probe beam is generated in the CALIFES linac and
accelerated to an energy of about 180~MeV before entering the Two-beam Test
Stand. In this experiment, 148~ns long probe beam bunch-trains were used with a
total charge of 10.2~nC or 0.1~nC per bunch. Because we observed slow
fluctuations of the energy and orbit of the probe beam over a few-minute time
scale, we need to consider relative measurements between accelerated and
non-accelerated bunch-trains which are subsequent in time. This can be achieved
in the Two-beam Test Stand operating the probe beam at twice the repetition rate
of the drive beam. We successfully operated the drive beam up to a repetition
rate of 2.5~Hz and the probe beam up to a repetition rate of 5~Hz. This way it
was possible to have a RF pulse sent to the accelerator structure every second
probe beam pulse and therefore to accelerate every other probe beam bunch-train,
with a non-accelerated bunch-train in between.

\section{\label{sec:accfx}Effect of the acceleration on the beam}

\begin{figure*}[p]
\subfigure[]{\label{fig:spot617}\includegraphics[width=0.45\linewidth]{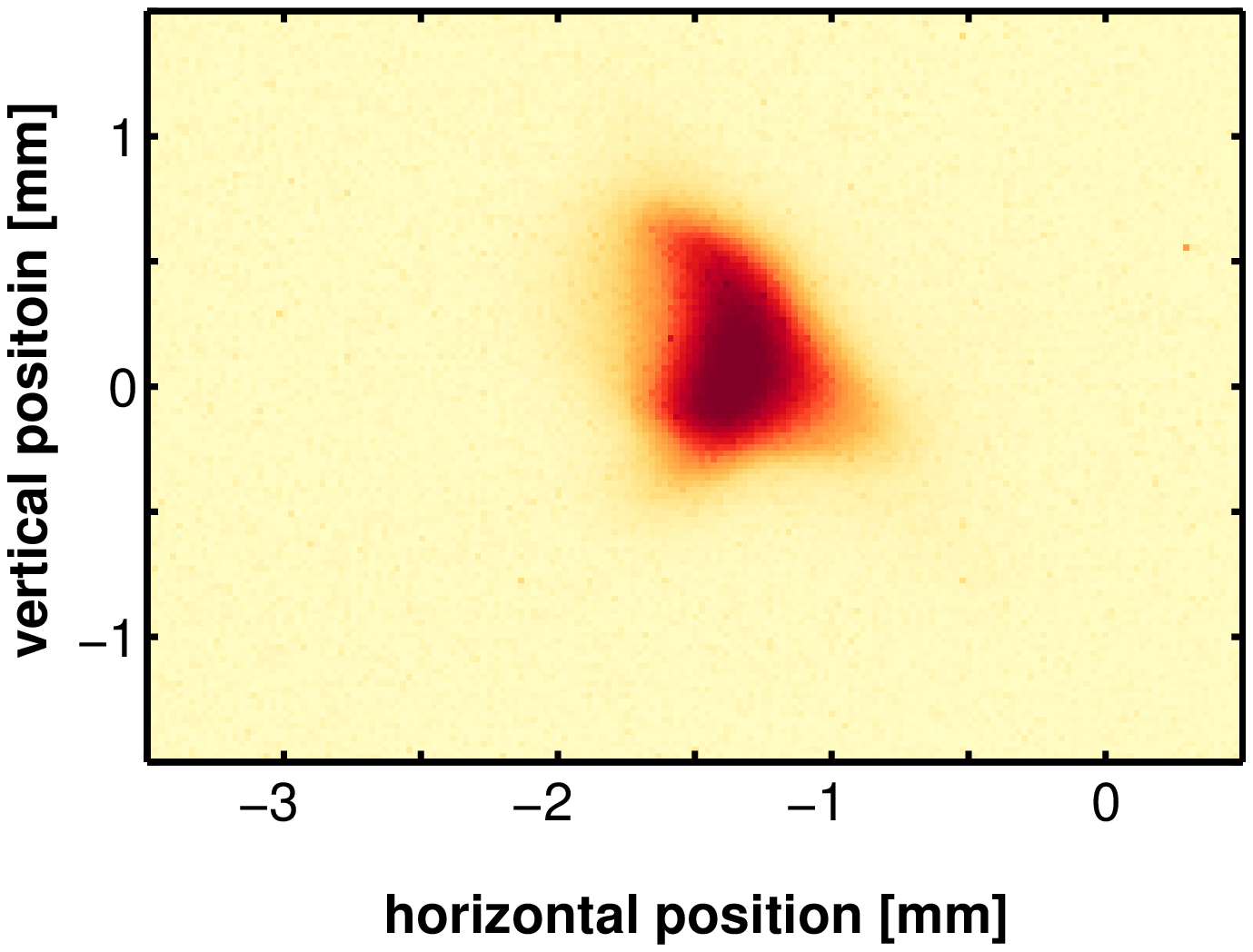}}
\subfigure[]{\label{fig:spot618}\includegraphics[width=0.45\linewidth]{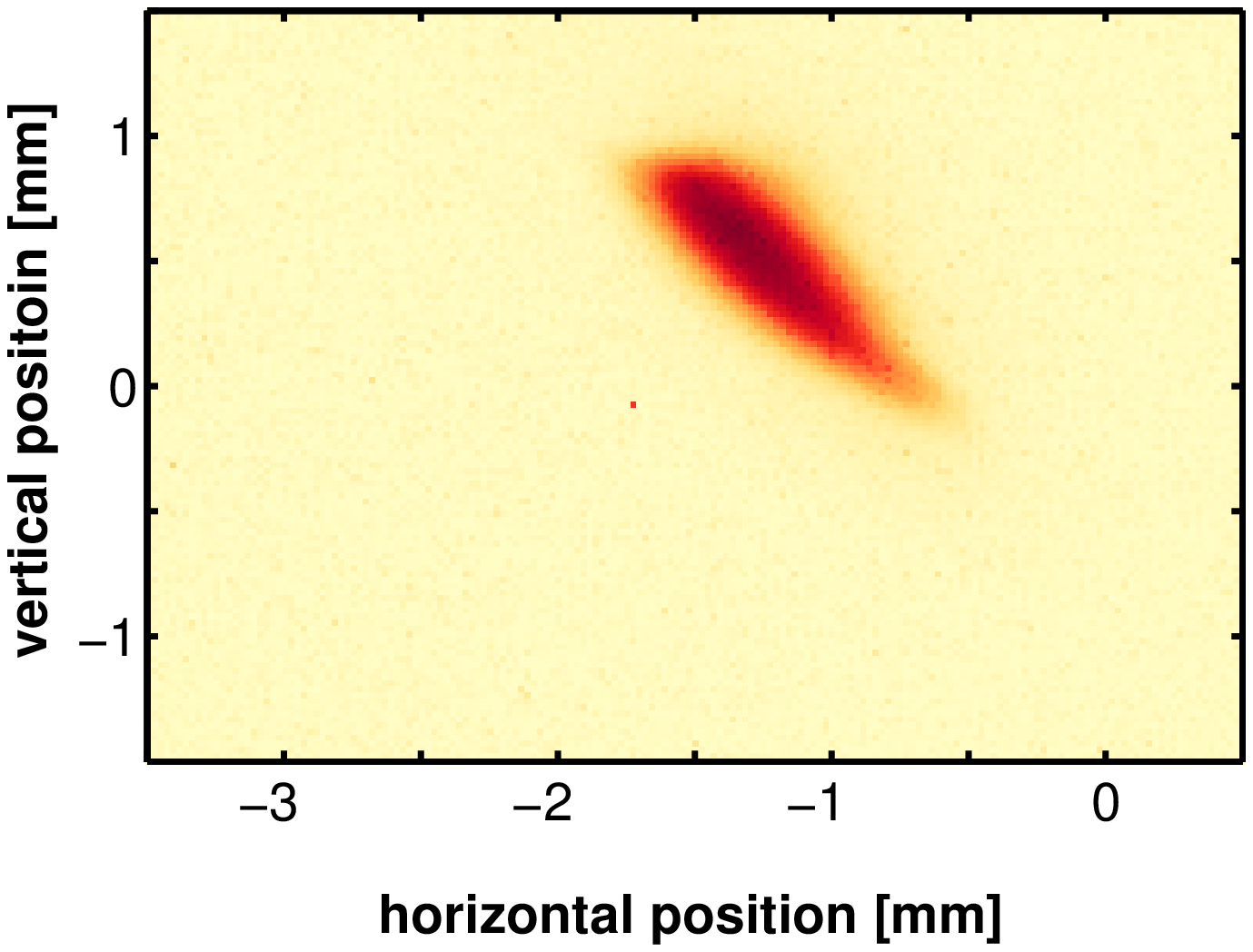}}
\subfigure[]{\label{fig:spot619}\includegraphics[width=0.45\linewidth]{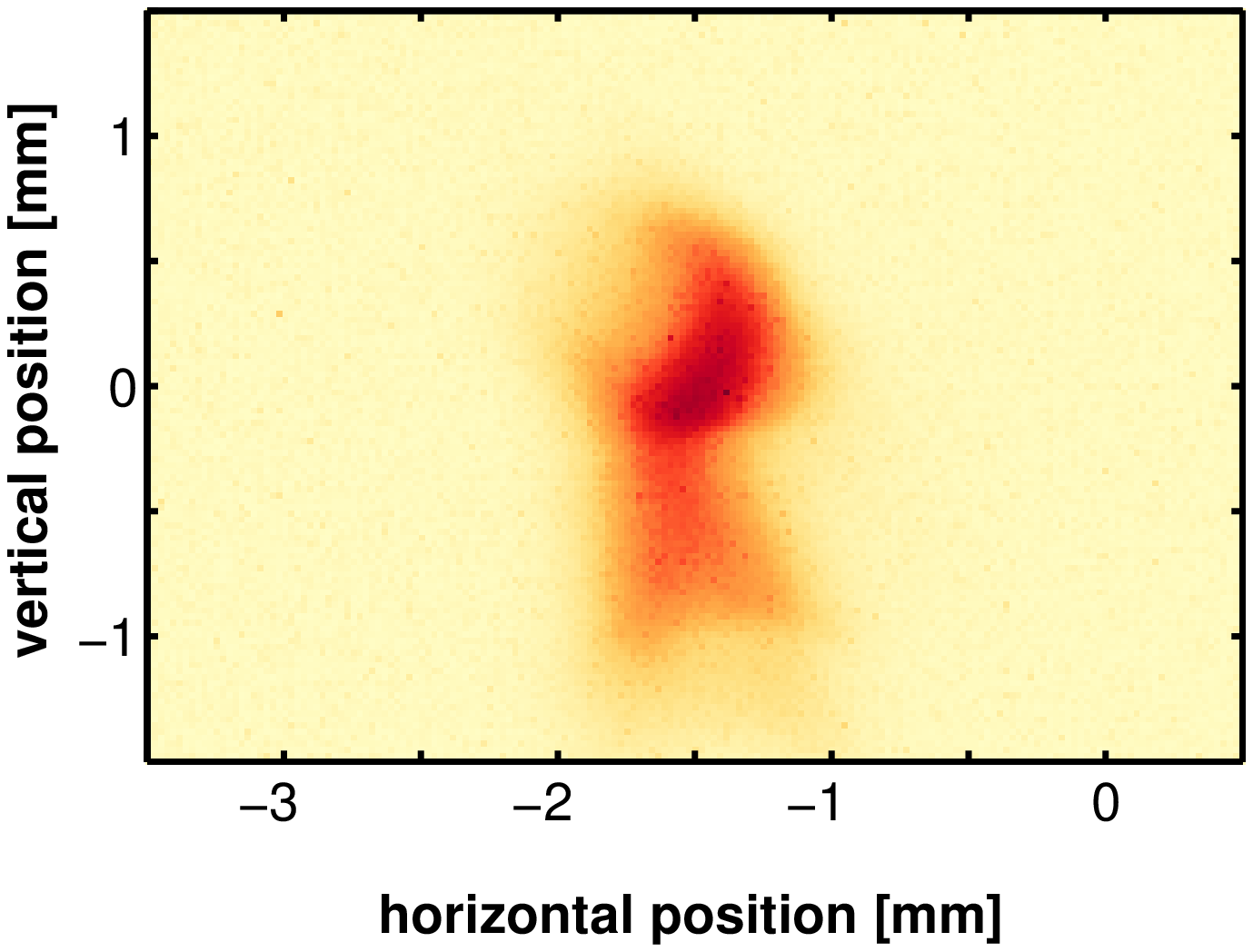}}
\subfigure[]{\label{fig:prediction619}\includegraphics[width=0.45\linewidth]{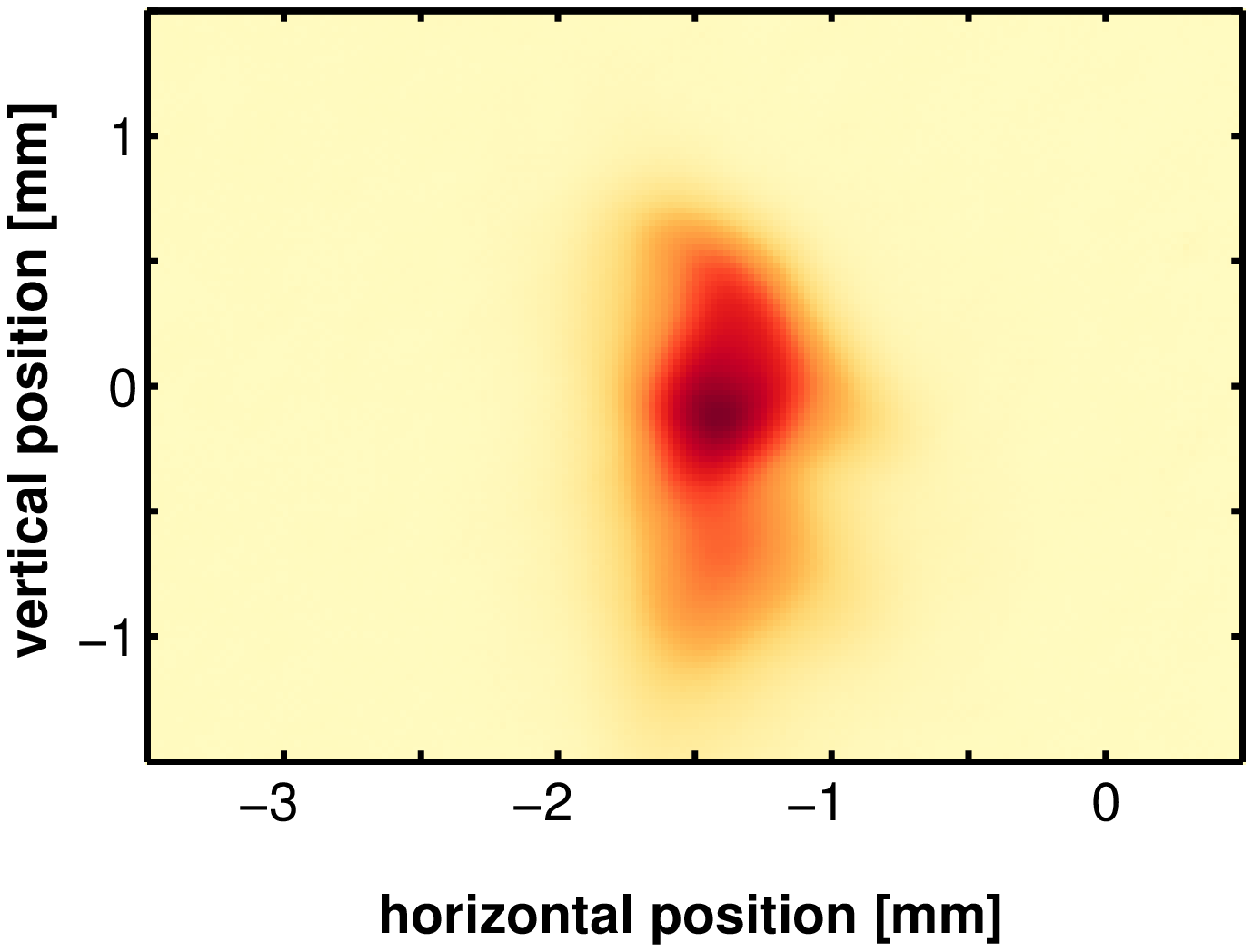}}
\subfigure[]{\label{fig:orbits_seq619}\includegraphics[width=0.45\linewidth]{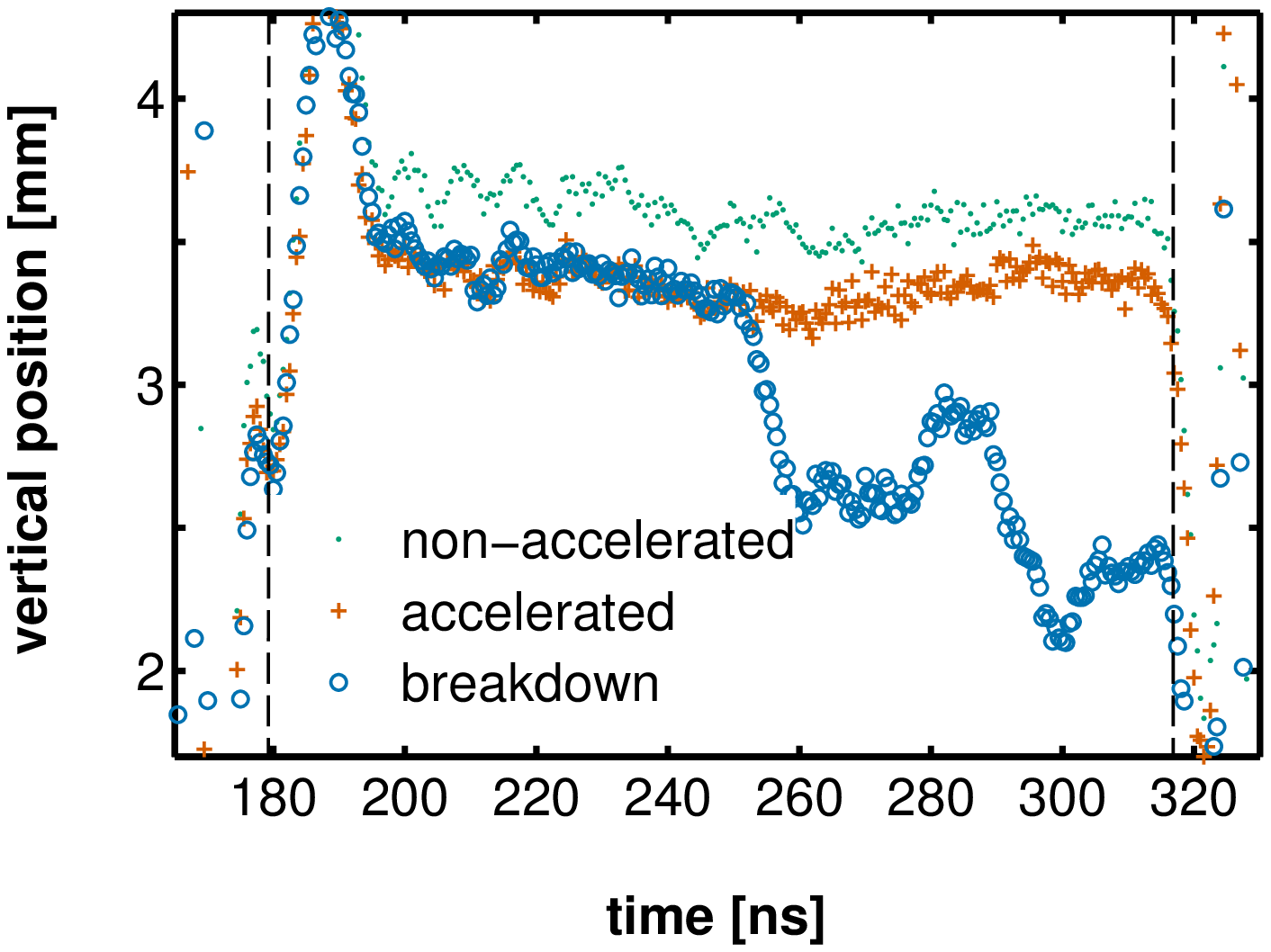}}
\subfigure[]{\label{fig:rf619}\includegraphics[width=0.45\linewidth]{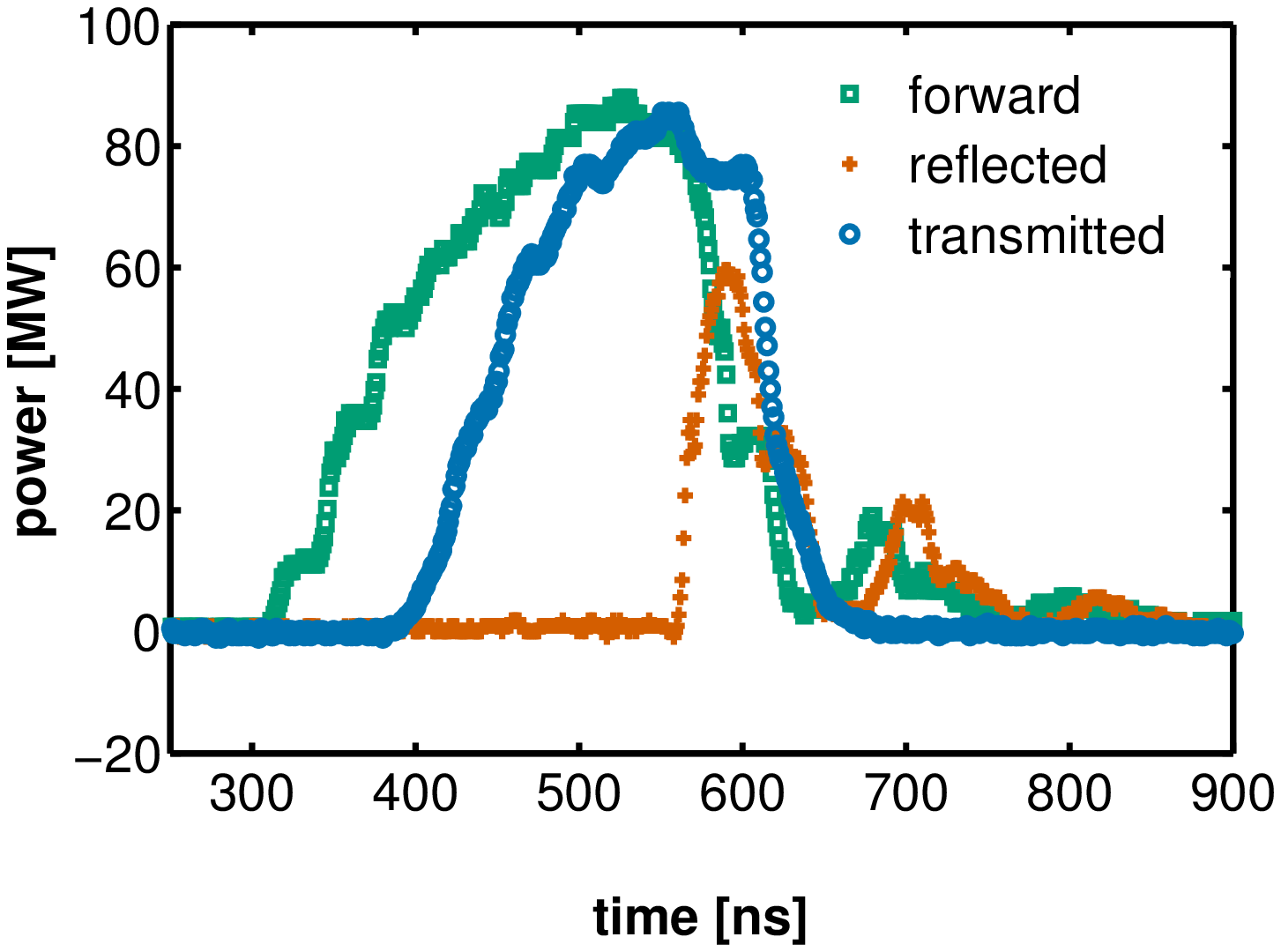}}
\caption{\label{fig:seq619}Collection of measurements from three consecutive
bunch-trains and prediction of the beam spot in case of RF breakdown: (a)
accelerated bunch-train; (b) non-accelerated bunch-train; (c) bunch-train
corresponding to a RF breakdown in the accelerator structure; (d) prediction of
the beam spot in case of RF breakdown; (e) beam orbit measurements; (f) power
measurements corresponding to a RF breakdown in the accelerator structure.}
\end{figure*}

Two representative examples of beam spots measured on the screen installed
before the spectrometer line (see Fig.~\ref{fig:probebeam}) are shown in
Fig.~\ref{fig:spot617} and Fig.~\ref{fig:spot618}. Both beam spots were measured
when neither the quadrupole triplet nor the correctors between accelerator
structure and screen were powered, i.e.\ the beam was following a ballistic
trajectory after the accelerator structure.
The beam spot in Fig.~\ref{fig:spot617} stems from a bunch-train accelerated in
the structure by about 23~MeV whereas the beam spot in Fig.~\ref{fig:spot618}
corresponds to a bunch-train non-accelerated. The size of the latter is
determined in the CALIFES linac and is about 0.4~mm in both planes throughout
this experiment. We observed that the centroids of the two beam spots have
different positions and shapes. The centroid position is calculated by means of
a fit of a 2D-Gaussian to the beam spot image and is found to differ by
$0.40\pm0.10$~mm between the two cases. Such difference was not due to any
changes in the incoming beam orbit as the beam centroid measured on the screen
was stable within $\pm$0.13~mm from pulse to pulse. In
Fig.~\ref{fig:orbits_seq619} we show the beam position measured by the cavity
beam position monitor installed just before the imaging screen. The two
upper-most signals correspond to the bunch-trains whose spot is shown in
Fig.~\ref{fig:spot617} and Fig.~\ref{fig:spot618}, the accelerated one at the
top and the non-accelerated one in the middle. The signal at the bottom
corresponds to a breakdown and is discussed in Sec.~\ref{sec:evidence4kicks}.
The accelerated and non-accelerated positions are about 0.26~mm apart on the
vertical plane whereas their shift is negligible on the horizontal plane. Given
the distances of the screen and the cavity beam position monitor from the
accelerator structure of 5~m and 3.4~m respectively, the different beam position
measured with the beam position monitor should correspond to a difference of
$0.26\cdot(5/3.5)=0.37$~mm between the position of the beam centroids measured
on the screen, which is in agreement with what was measured. In other words, the
beam was always kicked vertically by about 0.07~mrad whenever it was
accelerated.
Such kick corresponds to a momentum of about 13~keV/c for a beam energy of
180~MeV. Such effect is consistent with a misalignment of the accelerator
structure with respect to the rest of the beam line. Given a total energy gain
of 23~MeV in the 23~cm long accelerator structure, a tilt of
$0.013/23\simeq0.57$~mrad would correspond to a displacement of $(0.23/2) \cdot
0.57 = 66~\mu$m which is consistent with the mechanical accuracy of the
alignment of the structure in the beam line~\cite{priv:GRiddone}. It is worth
noting that even if the beam could be steered on an orbit which was stable
irrespective whether the beam was accelerated or not, we chose to maintain the
shift between accelerated and non-accelerated positions. That permits an
indirect measurement of the beam energy change, even if it introduces a bias in
the measurement of transverse breakdown kicks as discussed in
sec.~\ref{sec:evidence4kicks}.

Beyond the position of the centroids of the two beam spots in
Fig.~\ref{fig:spot617} and Fig.~\ref{fig:spot618}, we also observed that both
their shapes and orientations are different. First of all it is worth noting
that the beam spot corresponding to the non-accelerated bunch-train in
Fig.~\ref{fig:spot618} is elliptical and tilted, which indicates coupling
between the horizontal and the vertical plane. This coupling is caused by the
solenoids surrounding the CALIFES accelerator structures to provide transverse
focusing when the beam is generated and accelerated before being sent to the
Two-beam Test Stand. On the other hand, no solenoidal field is present in the
Two-beam Test Stand therefore we attribute the change in size and orientation of
the accelerated bunch-train to the field in the structure. If we consider that
the beam size of the accelerated beam spot measured on the screen is almost
doubled with respect to the non-accelerated one, such change can be accounted
for by a defocusing quadrupole with a focal length of 5~m.
This effect can be explained if we consider that a quadrupolar component of the
fundamental accelerating mode is present in both the input and output couplers
of the accelerating structure due to their symmetry which is broken by the
presence of two feeding waveguides. Moreover the symmetry of all the cells in
the accelerator structure is broken by the presence of wakefield damping
waveguides such that there an octupolar component of the fundamental mode is
introduced. The integrated strength of such additional multipolar modes were
calculated for a CLIC baseline accelerator structure~\cite{Grudiev:2012zzc} and
resulted in $17 \cdot 10^{-3}$~T integrated throughout the input and output
couplers and $8\cdot10^5$~Tm$^{-2}$ integrated throughout all regular cells of
the accelerator structure. For a beam energy of 180~MeV the quadrupolar field
component contributes with a focal length of about 30~m, which is not enough to
explain our observation. On the other hand, the slope of octupolar field
component in the regular cells at a beam offset of 200~$\mu$ m in the structure
produces a quadrupolar effect with a focal length of about 5~m. Thus the effect
of the octupolar mode in the regular cells of the accelerator structure can
account for the observed focusing effect.

\section{\label{sec:evidence4kicks}Evidence for breakdown effects on the beam}

When a RF breakdown occurred in the accelerator structure while the beam was
accelerated, we often observed a beam spot like the one shown in
Fig.~\ref{fig:spot619}. Such spot must be compared with the two immediately
preceding accelerated and non-accelerated beam spots in Fig.~\ref{fig:spot617}
and Fig.~\ref{fig:spot618}, respectively. A second spot is visible in
Fig.~\ref{fig:spot619} below the original spot, which we explain as resulting
from part of the bunch-train receiving a transverse momentum in correspondence
of the breakdown location in the structure, and therefore travelling
on a different orbit and hitting the screen in a different position. To verify that such a
secondary spot was part of the beam, we recorded images of discharge currents
reaching the screen without beam. This always resulted in low intensity images
of breakdown current spread over a wide portion of the screen. Therefore
discharge currents could be distinguished in every case from the beam and they
can be excluded as possible source of secondary spots detected in breakdown
events.
The explanation offered for the appearance of a second beam spot is supported by
the measurements of the beam position on a nano-second time scale with the
cavity beam position monitor installed just before the imaging screen, as shown
in Fig.~\ref{fig:orbits_seq619}. The position of the first part of the
bunch-train - from 180~ns to 250~ns - is the same of the position of an
accelerated bunch-train when no breakdown happens in the accelerator structure,
i.e.\ it gets fully accelerated.
Afterwards the position shifts in about 10~ns vertically downwards by about
0.75~mm. It oscillates for the remaining length of the bunch-train but it never
shifts back to its initial position. It is worth noting that the overshoot
visible at the beginning of the signal is due to the read-out electronics and
that it is not related to the beam position. Moreover it is a feature observed
in all the signals read-out from the cavity beam position monitors in the beam
line. Therefore if we disregard the first 20~ns of such signals, we can identify
two main trajectories, the first one corresponding to the fully accelerated beam
and the second one corresponding to the beam affected by the breakdown. These
two positions will result in two distinct spots on the imaging screen. The
position of such spots is calculated as the distance between the two maxima of
the sum of two 2D-Gaussian fitted to the beam image and is 1.1~mm in
Fig.~\ref{fig:spot619}.
Given the distance of the screen and the cavity beam position monitor from the
accelerator structure, the distance between the two spots measured on the screen
corresponds to the shift of the beam position measured by the beam position
monitor in Fig.~\ref{fig:orbits_seq619} and it is $1.1\cdot(3.4/5)=0.68$~mm.

A further indication that the rapid deviation of the beam position measured on
the screen and with a beam position monitor is correlated with a RF breakdown in
the accelerator structure is given by the RF measurement shown in
Fig.~\ref{fig:rf619}. Such measurements, corrected for the ohmic losses in the
accelerator structure and the waveguides, show that almost 100\% of the power
fed into the accelerator structure is reflected backwards. The time at which the
reflection is measured is consistent with the time at which the beam position
recorded by the beam position monitor starts changing.

\section{\label{sec:scenarios}Breakdown scenarios}

\begin{figure*}[p]
\subfigure[]{\label{fig:spot590}\includegraphics[width=0.45\linewidth]{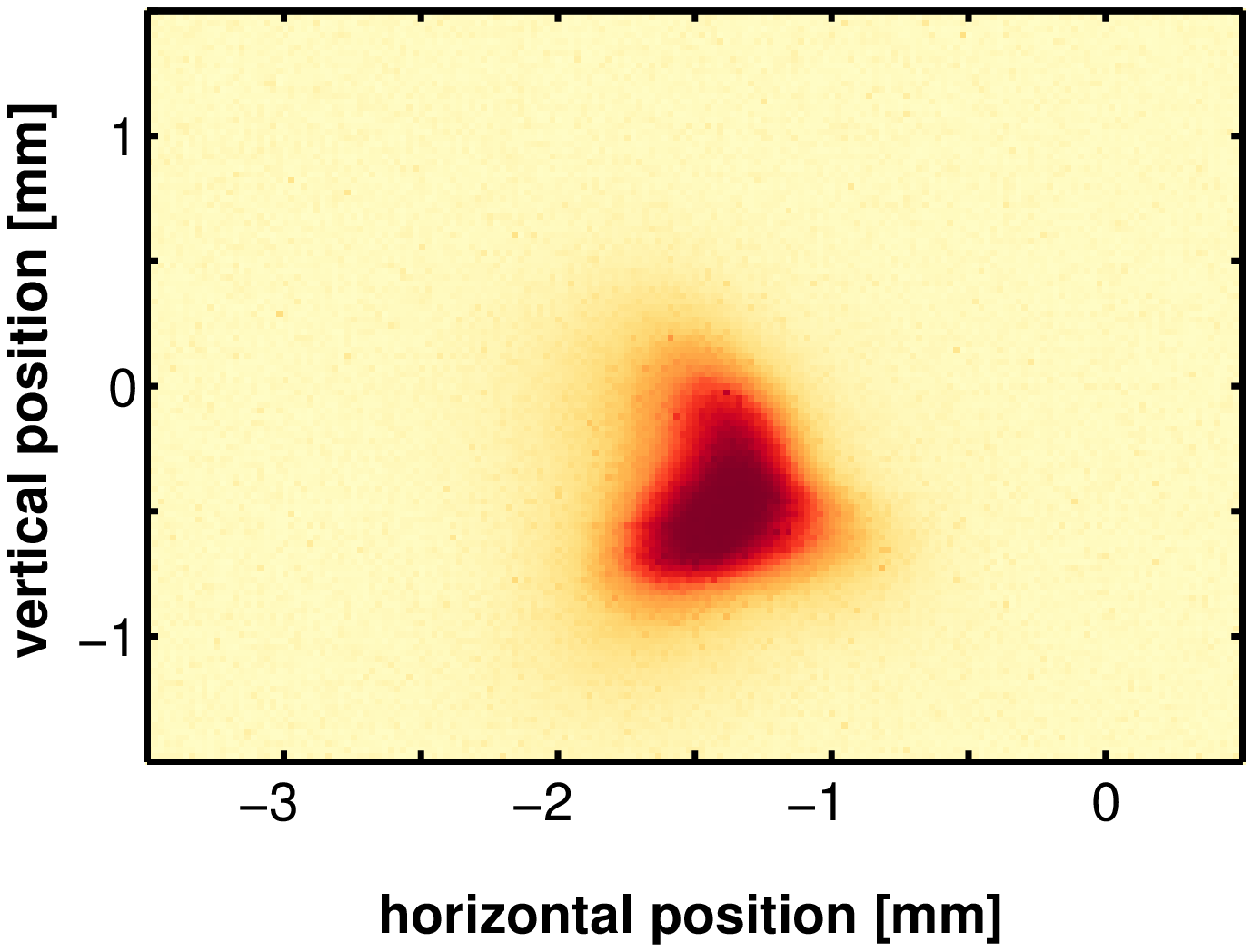}}
\subfigure[]{\label{fig:spot591}\includegraphics[width=0.45\linewidth]{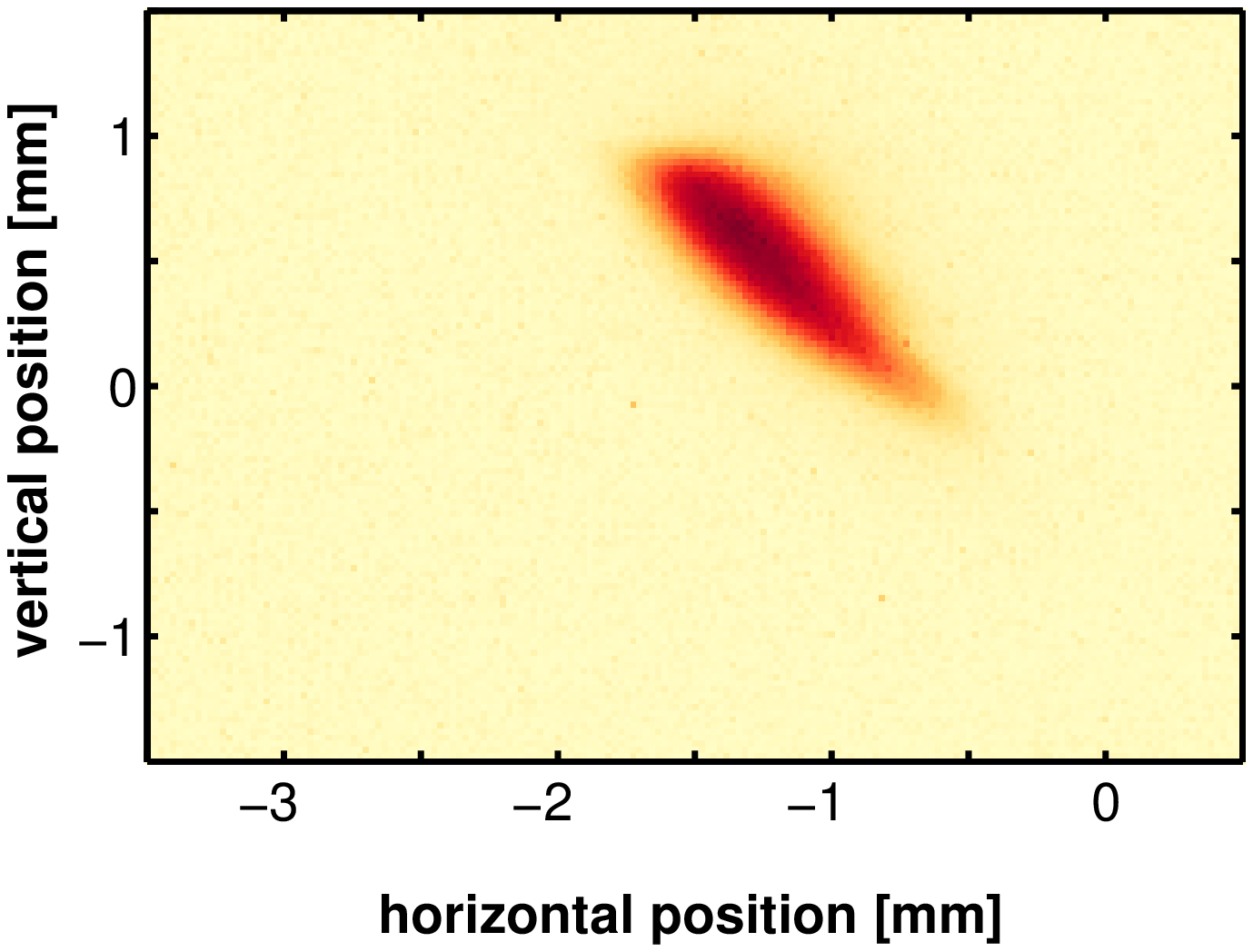}}
\subfigure[]{\label{fig:spot592}\includegraphics[width=0.45\linewidth]{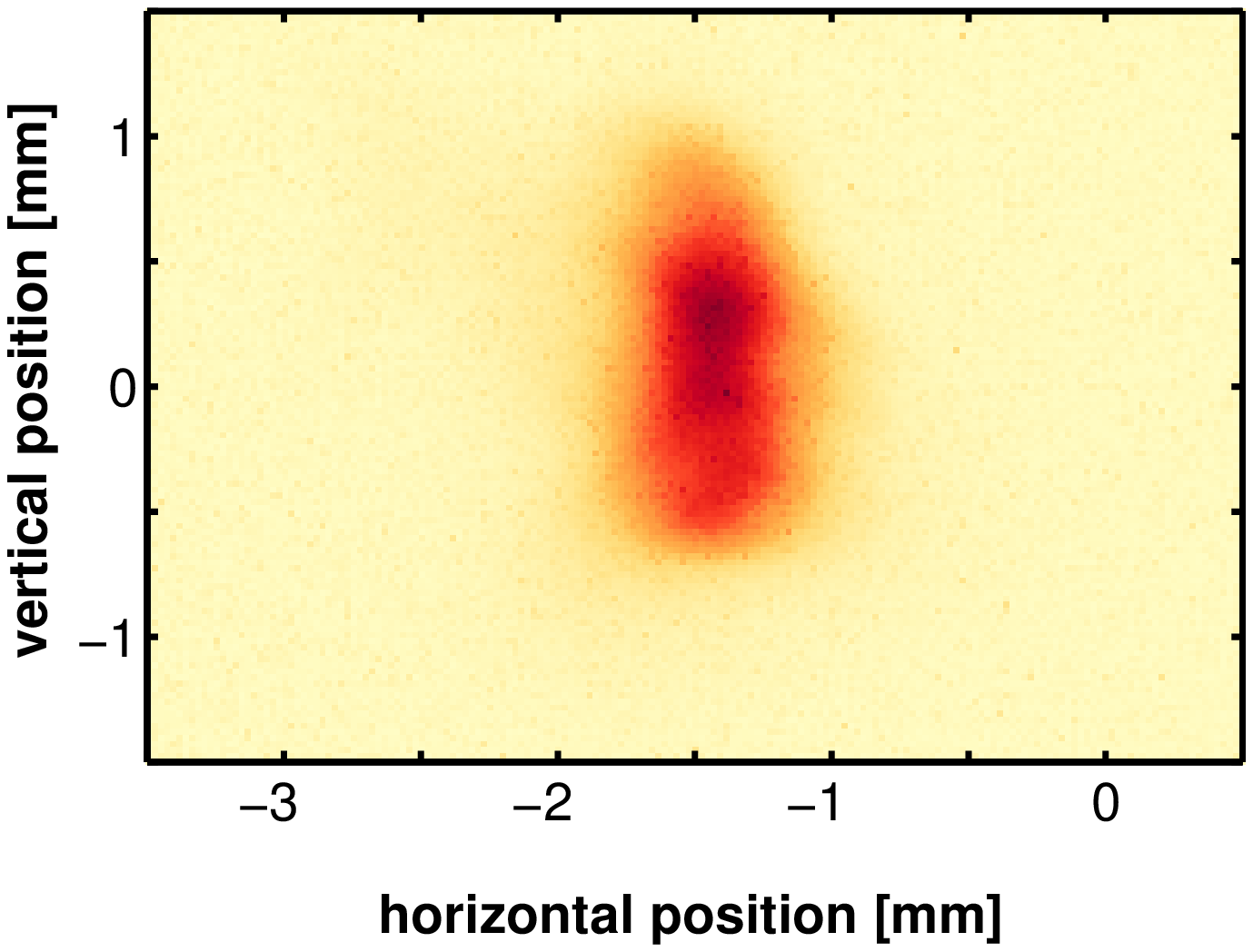}}
\subfigure[]{\label{fig:prediction592}\includegraphics[width=0.45\linewidth]{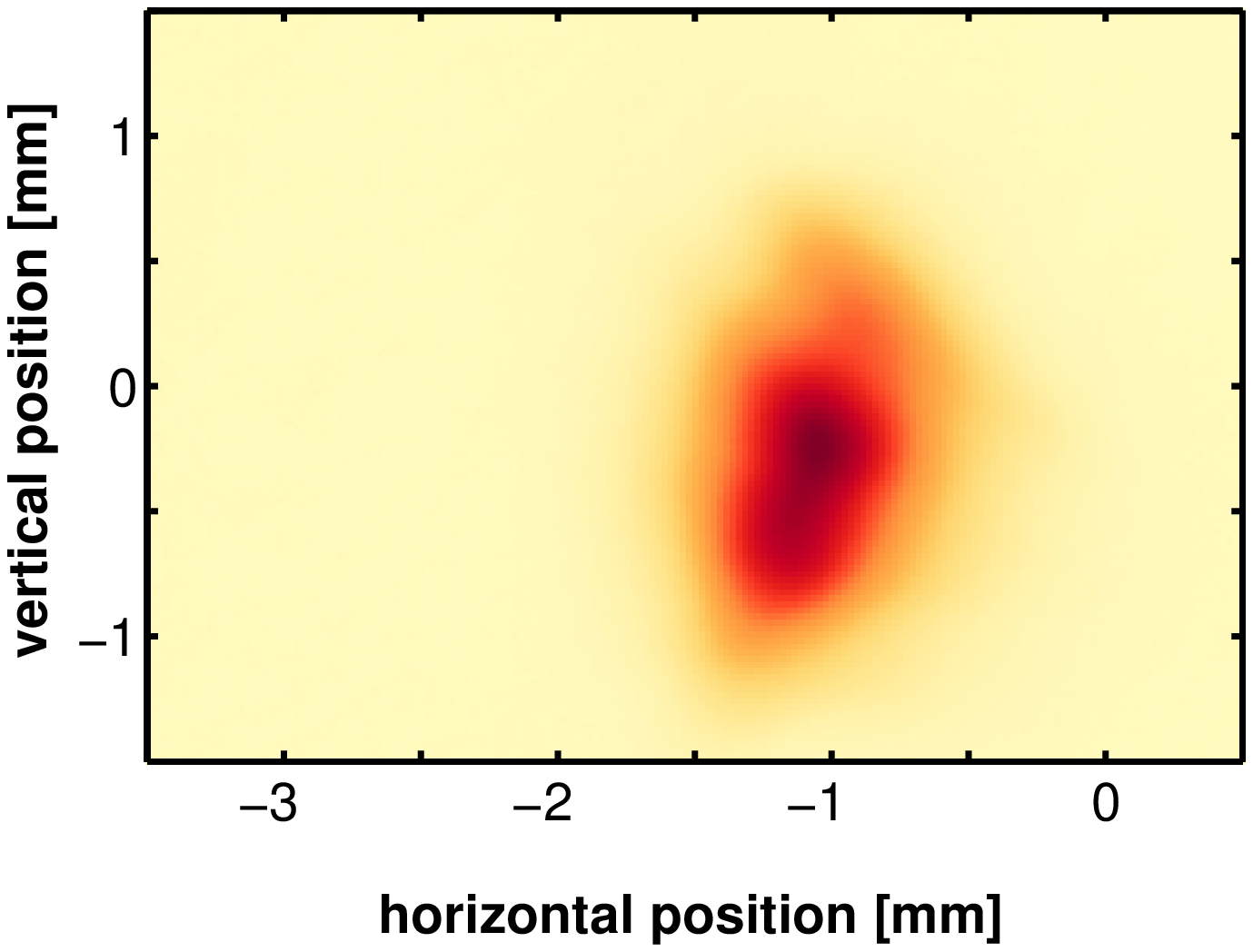}}
\subfigure[]{\label{fig:orbits_seq592}\includegraphics[width=0.45\linewidth]{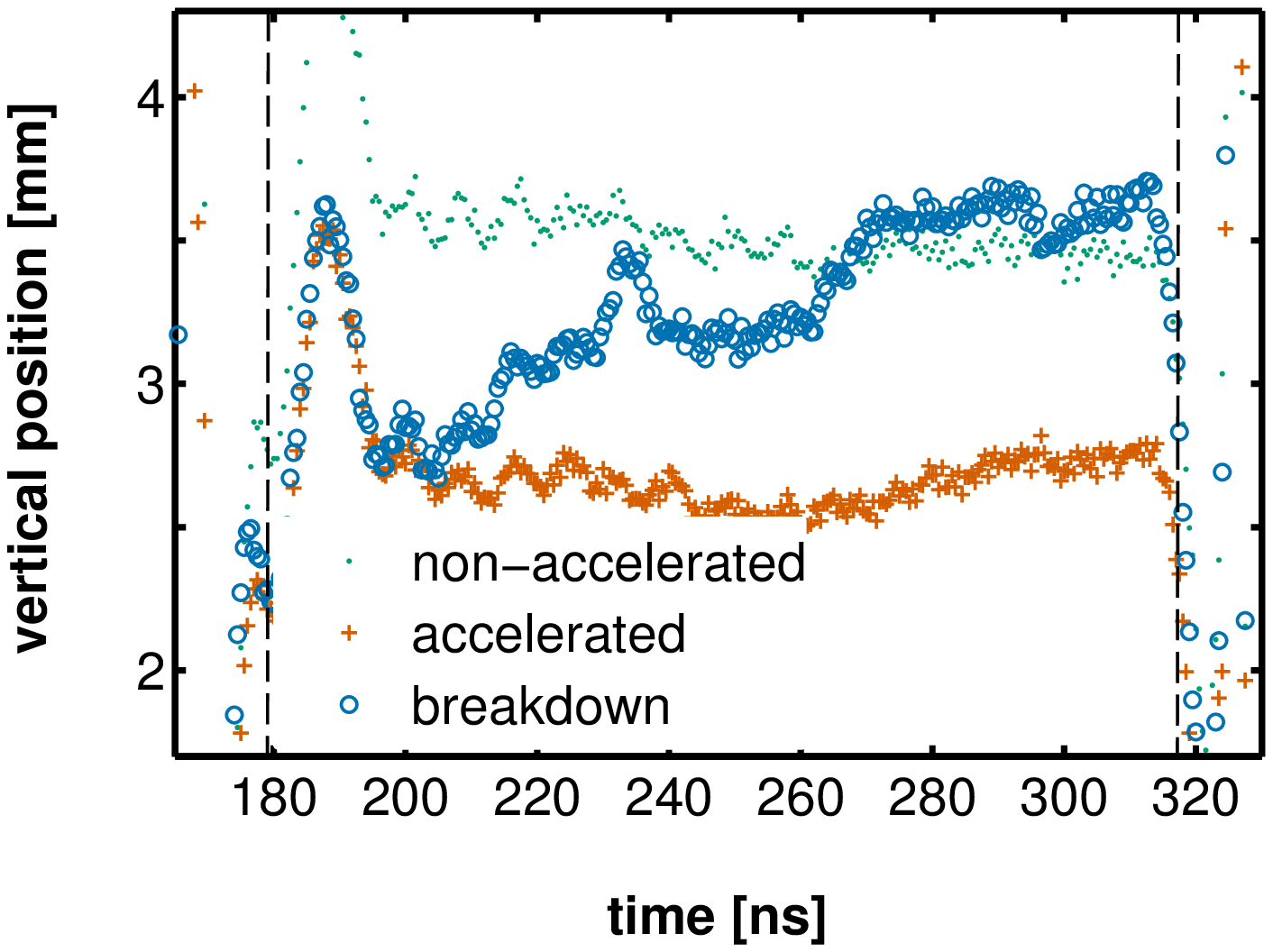}}
\subfigure[]{\label{fig:rf592}\includegraphics[width=0.45\linewidth]{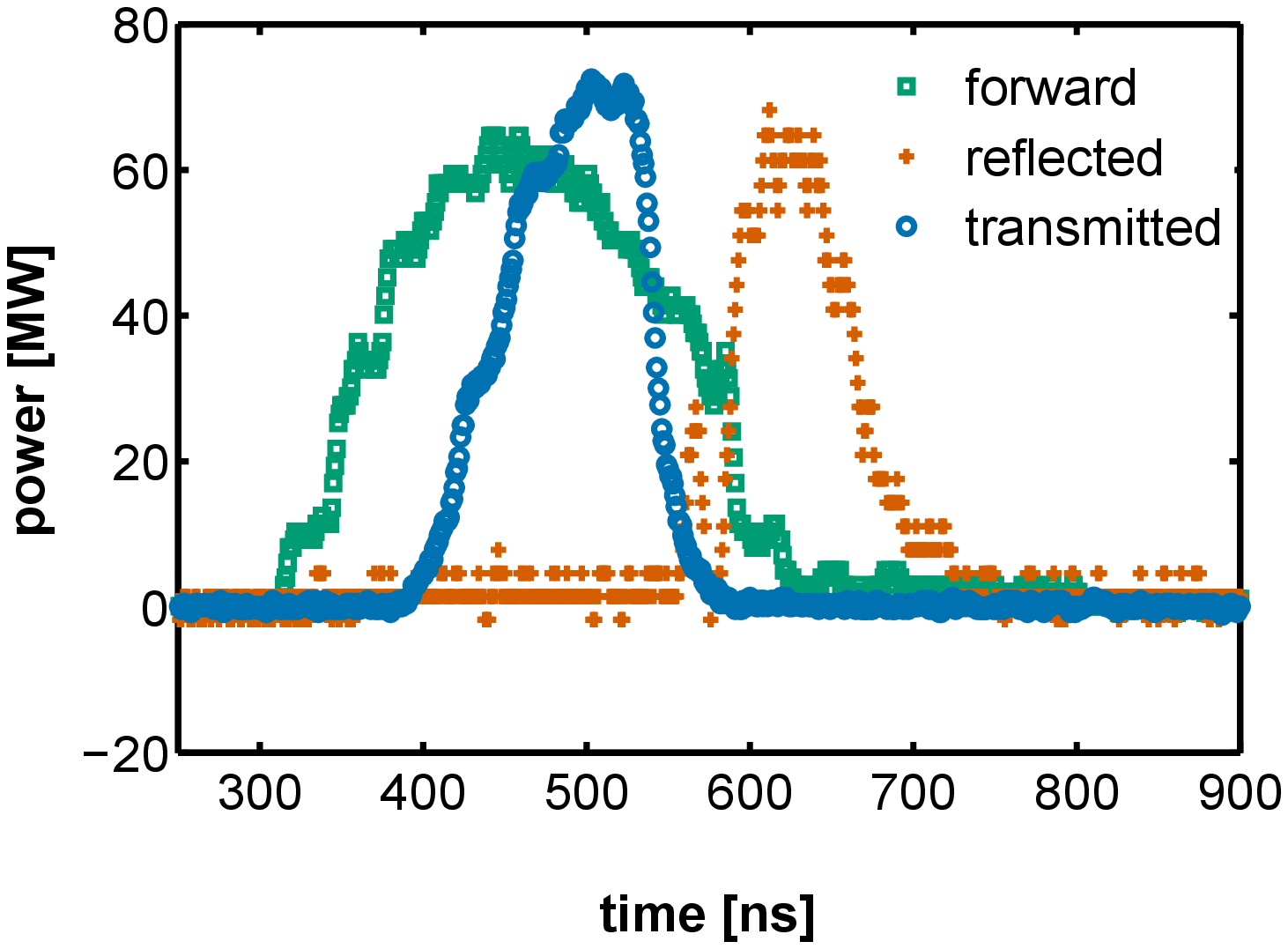}}
\caption{\label{fig:seq592}Collection of measurements from three consecutive
bunch-trains and prediction of the beam spot in case of RF breakdown: (a)
accelerated bunch-train; (b) non-accelerated bunch-train; (c) bunch-train
corresponding to a RF breakdown in the accelerator structure; (d) prediction of
the beam spot in case of RF breakdown; (e) beam position measurements; (f) power
measurements corresponding to a RF breakdown in the accelerator structure.}
\end{figure*}

The RF breakdowns measured during this experiment have different characteristics
and in addition to the breakdown in Fig.~\ref{fig:seq619} discussed in
sec.~\ref{sec:evidence4kicks}, we discuss here two particular cases to
illustrate other breakdown scenarios. The plot in Fig.~\ref{fig:orbits_seq592}
shows a case in which the beam is accelerated for about 20~ns until its vertical
position drifts upwards over about 65~ns and finally remains consistent with the
position of a non-accelerated bunch-train.
The 65~ns drift time of the accelerated position towards the non-accelerated
corresponds to the filling time of the accelerator structure. This suggests the
following explanation: the first 20~ns of the bunch-train are fully accelerated
and then a breakdown happens close to the input port of the accelerator
structure. From that moment on, most of the power fed to the structure is
reflected backwards and therefore the power which fills the accelerator
structure flows towards its output port in 65~ns until the accelerator structure
is left almost completely empty. The beam which in the meantime is still
travelling in the accelerator structure is less and less accelerated therefore
its position slowly drifts towards the position of a non-accelerated
bunch-train. Because of the small group velocity in the accelerator structure -
from 1.6\% to 0.8\% of the speed of light - we can estimate the location of the
breakdown in the accelerator structure by comparing power measurements at the
input and output directional couplers (see Fig.~\ref{fig:probebeam}). For the
breakdown event in Fig.~\ref{fig:seq592} we estimated the breakdown to be
localised in the sixth cell of the structure, therefore close to its input port.
Moreover the RF measurements for this event in Fig.~\ref{fig:rf592} show that
almost 100\% of the RF fed to the accelerator structure is reflected backwards
and that the amount of power that leaks through the breakdown location drops in
about 10~ns down to 50\% of the input power, and then down to zero during the
following 50~ns.
Already at 50\% of the input power the acceleration is reduced by a factor
$\sqrt{2}$ which is consistent with the drift of the beam position from the
accelerated to the non accelerated one as shown by the beam position monitor
signal in Fig.~\ref{fig:orbits_seq592}.

\begin{figure}[htb]
  \includegraphics[width=0.9\linewidth]{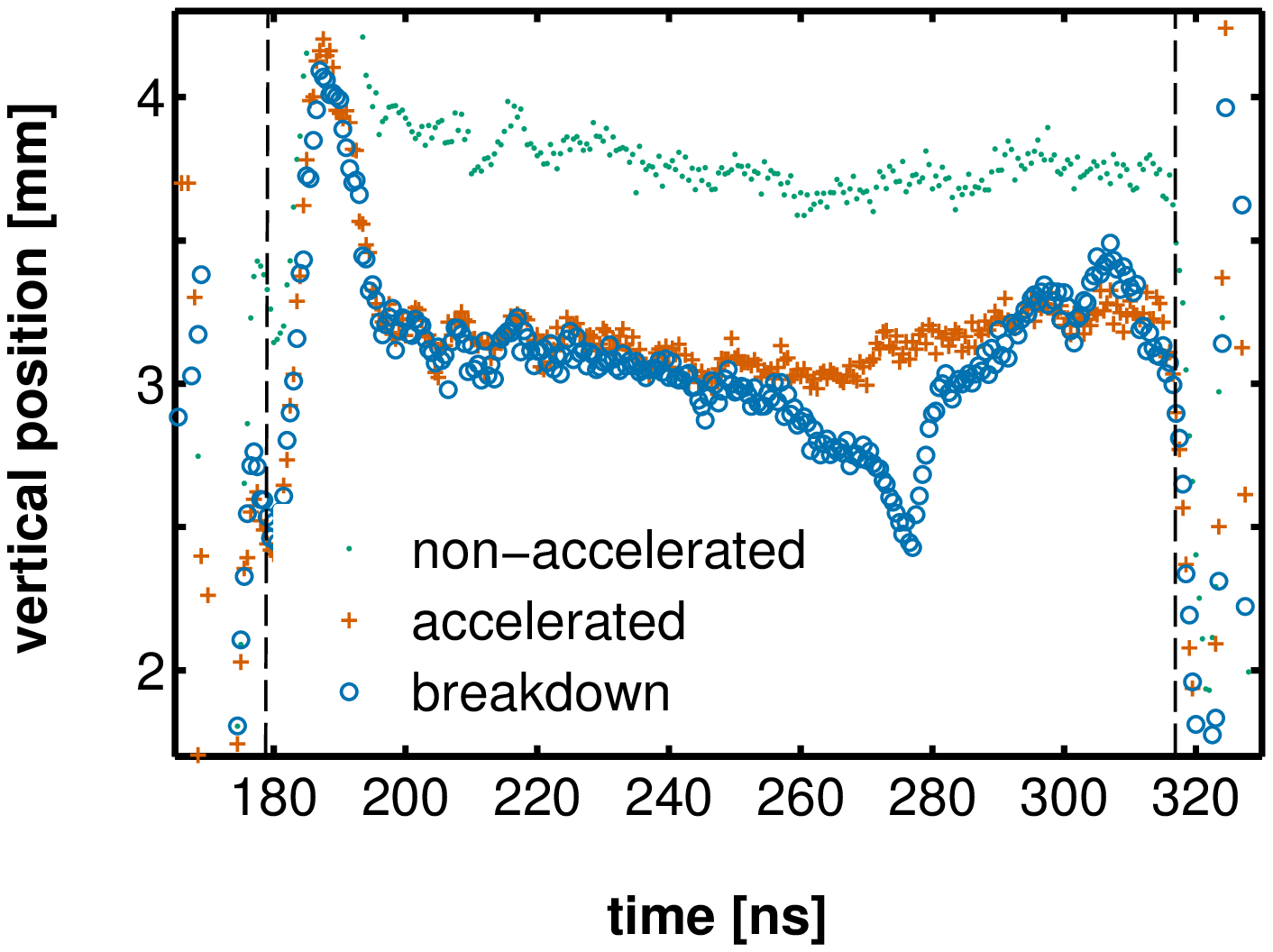}
  \caption{\label{fig:BDscenario2}Vertical beam position measurements corresponding
  to a non-accelerated bunch-train at the top, an accelerated one in the middle
  and one corresponding to a RF breakdown during the acceleration at the bottom.
  The bleep in the bottom signal corresponds to the effects of a short breakdown
  that displaces downwards the beam position.}
\end{figure}

A second breakdown scenario is shown in Fig.~\ref{fig:BDscenario2} where the
beam position measured with a cavity beam position monitor has a dip in
correspondence of a breakdown which lasts for about 20~ns. In this case, during
the breakdown the beam is pushed away from the non-accelerated one as it was the
case for the breakdown shown in Fig.~\ref{fig:seq592}, suggesting that little or
no contribution comes from lack of accelerating gradient after the breakdown
location or that the breakdown happened towards the output port of the
accelerator structure. The estimation of the breakdown location based on RF
measurements supports indeed the latter hypothesis, i.e.\ that the breakdown
happened in the last cell of the accelerator structure.

To test the explanation that we offered to describe a breakdown in the
accelerator structure and its effect on the beam orbit and energy, we try to
predict how the beam spot looks like on the imaging screen for the breakdown
shown in Fig.~\ref{fig:spot619} and Fig.~\ref{fig:spot592}.
According to what discussed above, we expect the breakdown beam spot to be a
mixture of both the non-accelerated and the accelerated beam spots. When the
bunch-train is fully accelerated in the accelerator structure its spot
corresponds to an unperturbed accelerated beam spot, whereas after a breakdown
we expect it to be a mixture of accelerated and non-accelerated spots according
to the proximity of the beam to the accelerated or non-accelerated positions,
respectively.
In other words, the breakdown spot can be expressed by a linear combination of
the accelerated and of the non-accelerated beam spots, whose coefficients at a
given time along the bunch-train are derived from the beam position monitor
measurements, according to the beam position. The results obtained are shown in
Fig.~\ref{fig:prediction619} Fig.~\ref{fig:prediction592}.
We want to stress that the displacement of the beam position in case of a RF
breakdown can be due to two different effects, the lack of power in the
structure due to its reflection and the direct effect of the breakdown on the
beam. When both affect the beam position in the same direction the two effects
cannot be disentangled unless we use the beam position monitor in the
spectrometer line, which is obscured by the screen used in our measurements.

\section{\label{sec:stat}Statistics}

\begin{figure}
 \subfigure[]{\label{fig:gradient_distro}\includegraphics[width=0.85\linewidth]{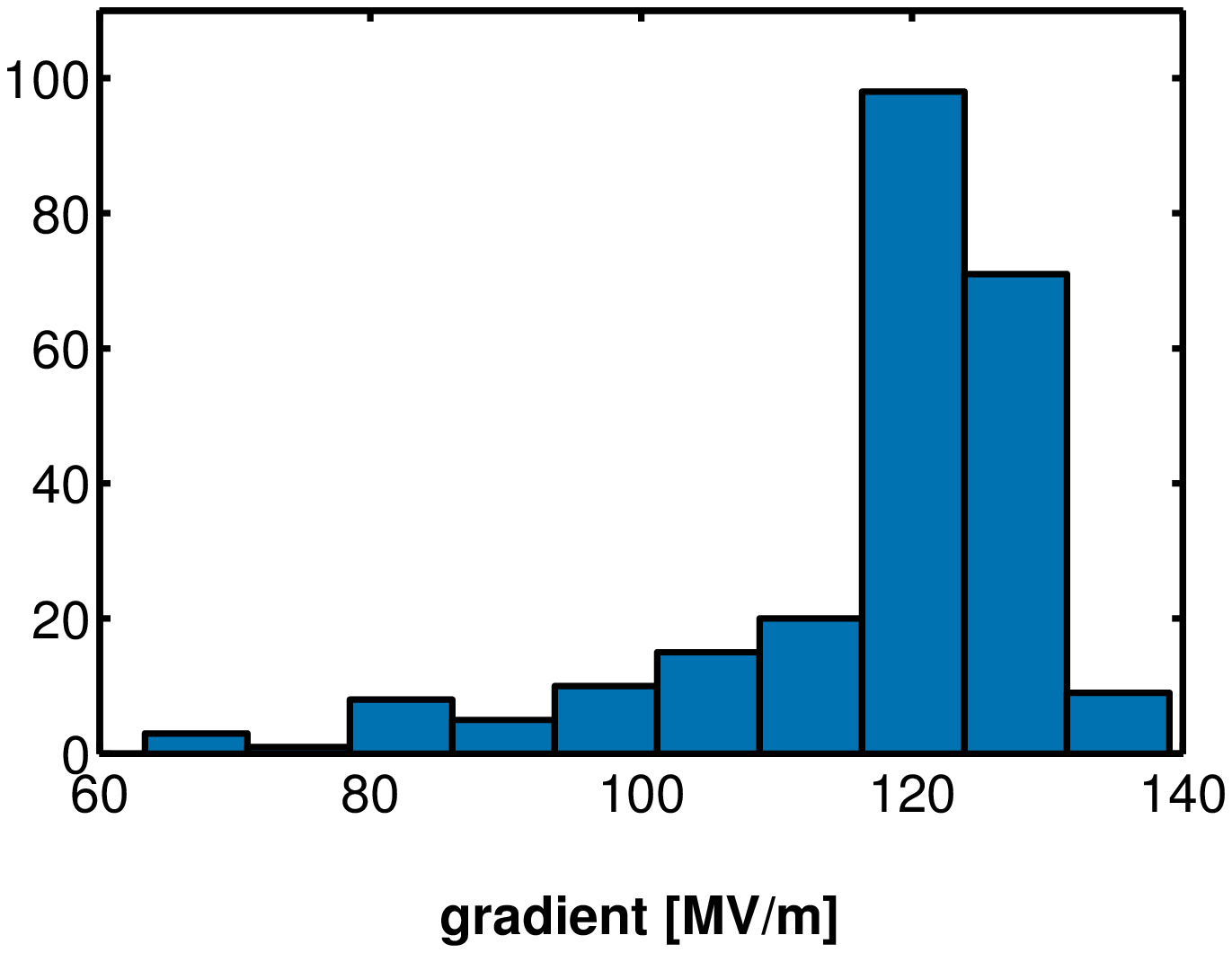}}
 \subfigure[]{\label{fig:misseDistro}\includegraphics[width=0.85\linewidth]{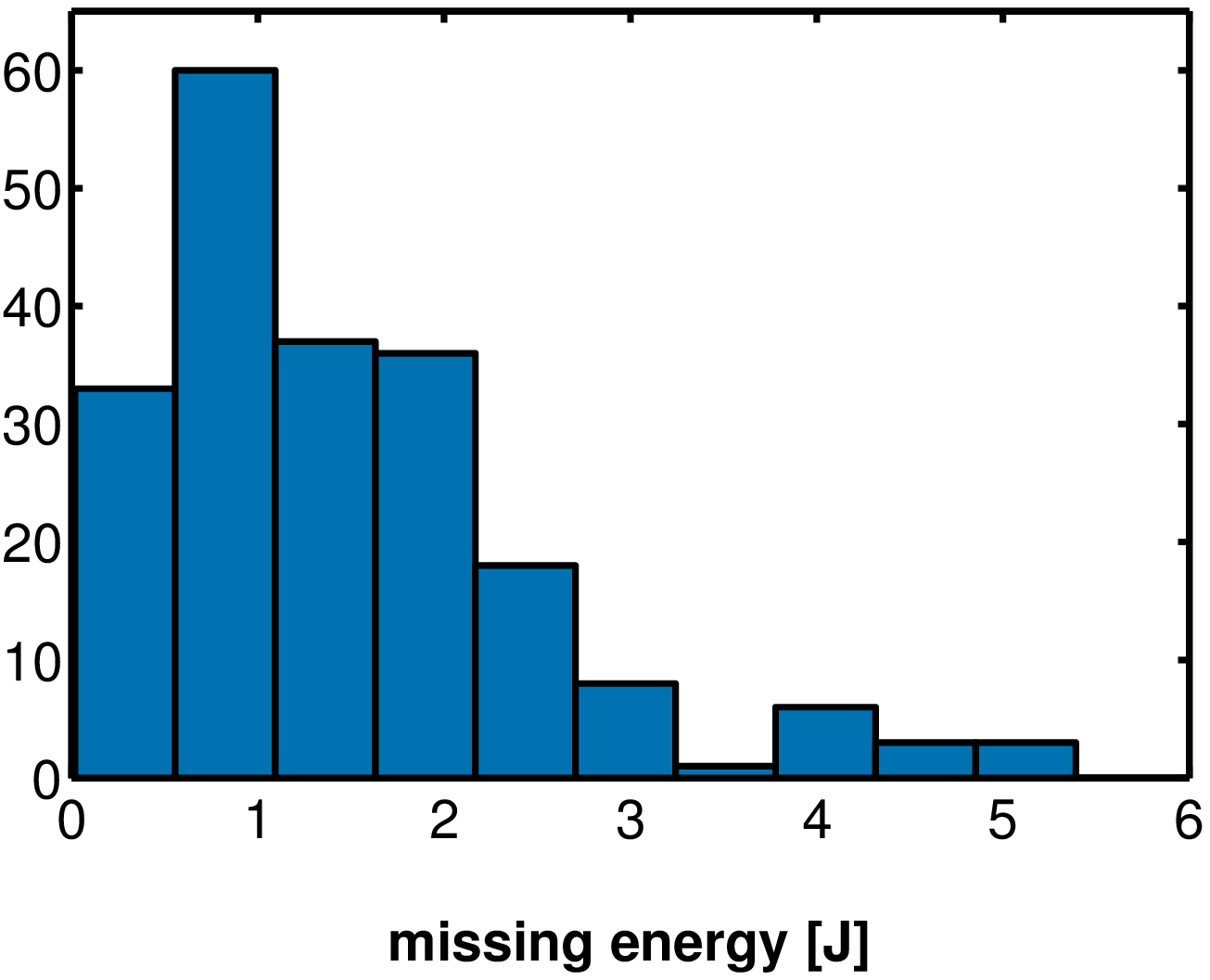}}
 \caption{\label{fig:bdstat}Distribution of (a) accelerating gradient and (b)
 missing energy for the 246 breakdown events considered in our
 analysis.}
\end{figure}

\begin{figure}
  \includegraphics[width=0.85\linewidth]{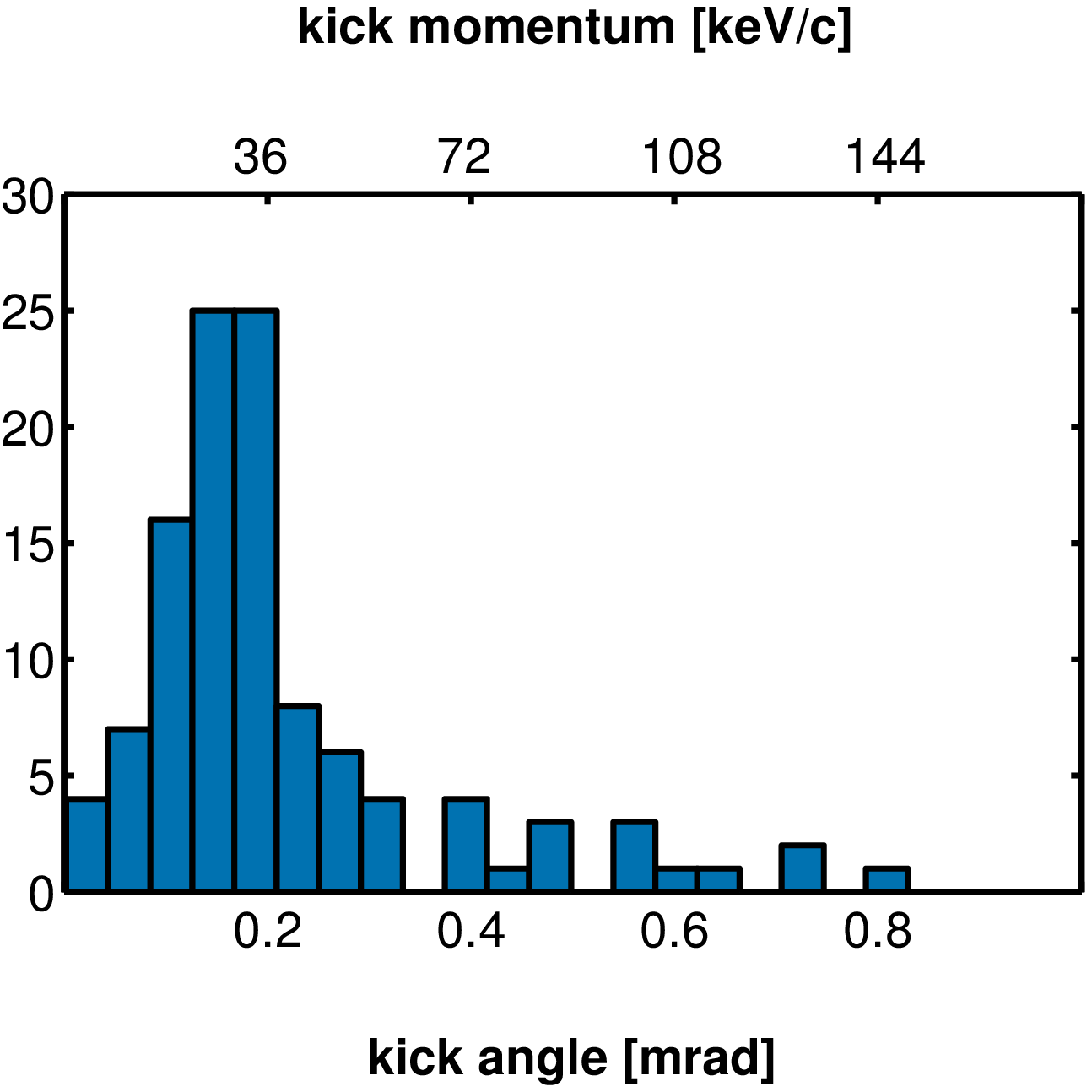}
  \caption{\label{fig:kickAngle_distro}Distribution of the total magnitude of RF
 breakdown kicks to the beam orbit.}
\end{figure}

The analysis described in sec.~\ref{sec:evidence4kicks} and
sec.~\ref{sec:scenarios} for three selected breakdowns has been applied to a
bigger data set of 246 breakdowns. Accelerating gradient and missing energy
calculated for those events are shown in Fig.~\ref{fig:bdstat}.
The accelerating gradient is calculated from the measurement of RF power fed to
the accelerator structure, given that the nominal power required to have an
average unloaded gradient of 100 MV/m is 46.5 MW (see
Table~\ref{tab:td24_params}). The energy absorbed in each breakdown is
calculated as the difference between the energy fed to the structure and the
energy transmitted to the structure output plus the energy reflected to the
structure input. Attenuation due to ohmic losses is taken into account.
The histogram in Fig.~\ref{fig:kickAngle_distro} shows the distribution of the
total angular magnitude of transverse kicks to the beam orbit, whose average
magnitude is $0.16\pm0.08$~mrad or $29\pm14$~keV/c.
As discussed in sec.\ref{sec:scenarios}, this estimation is biased by the effect
of the transverse beam acceleration due to the misalignment of the accelerator
structure and the beam orbit. Nevertheless, from the analysis of only the non
breakdown events we deduce that the maximum magnitude of such effect is about
one third of the average magnitude of the kick distribution presented in
Fig.~\ref{fig:kickAngle_distro}. In other words the distribution in
Fig.~\ref{fig:kickAngle_distro} represents mainly the effect of breakdown kicks,
because the contribution of acceleration kicks is smaller than 30\%.
Finally, we looked for correlations between the magnitude of the kick angle and
the location of the breakdown in the accelerator structure or the energy
dissipated in the breakdown. None of these correlations is statistically
significant.

\section{Conclusions}

We measured evidence of the effect of RF breakdown on the Two-beam test Stand
probe beam accelerated in a CLIC prototype structure. The position of
bunch-trains accelerated in the presence of a breakdown in the accelerator
structure, showed transverse displacement which we explain as resulting from
part of the bunch-train receiving a transverse momentum in correspondence of the
breakdown location, and therefore travelling on a different orbit. Both parts of
such bunch-trains were also intercepted and measured on a scintillating screen
where they resulted in two distinct spots whose distance is compatible with the
position measured with the beam position monitors in the beam line. We refer to
this transfer of transverse momentum to the beam as RF breakdown kick and we
observed that it does not depend on the amount of power dissipated by a
breakdown nor on the location at which a breakdown takes place.
We measured and discussed different examples of such kicks which suggest
different scenarios according to specific breakdown strength, duration and
location in the accelerator structure.

Furthermore we observed another source of transverse kicks to the beam orbit
which overlaps with the kicks given by breakdown. Such kick was found to be
accountable to a misalignment between the beam orbit and the axis of the
accelerator structure, and it always had a well defined magnitude and direction.
It represents a bias in the measurements of breakdown kicks though, which was
not possible to subtract. Nevertheless it provided an indirect measurement of
the beam energy during a breakdown, which we did not measure directly due to the
low resolution of the beam position monitor in the spectrometer line.

The average magnitude of transverse kicks to the beam measured in this
experiment is of $29\pm14$~keV/c in terms of transverse momentum transferred to
the beam, or $0.16\pm0.08$~mrad in terms of the angle given to the 180~MeV probe
beam. We consider as worst case scenario the case in which only RF breakdown
contribute to the measured kicks because such kicks cannot be avoided.
The worst place where a breakdown kick can happen in CLIC is at the beginning of
the main linac where the beam energy is 9~GeV. For such energy a transverse kick
of 29~keV/c corresponds to a kick angle of about 4~$\mu$rad. It is worth noting
that such angle is one order of magnitude bigger than the nominal CLIC beam
divergence of about 0.3~$\mu$rad, assuming a vertical normalised emittance of
10~nm mrad and a beta function of 10~m at the beginning of the main linac.

On the basis of our observations we suggested an explanation of how the beam
energy is affected by a breakdown, due to the reflection of power at the
breakdown location and the resulting change of the net acceleration in the
accelerator structure. We offered a model to predict how the beam spot looks
like in case of a breakdown, starting from measurements of breakdown beam
position and non-breakdown beam spots. The result approximates the measurement
although a more complete model which takes into account the phase of the
reflected power might help achieving increased agreement with the measurements.

Our interpretation of the observed kicks are magnetic fields generated by high
currents produced in the structure during a breakdown. In order to estimate the
order of magnitude of the currents and fields we assume that an arc is
established between two adjacent irises, where the electric fields assume their
extreme values. According to Ampere's law, a current of 250~A causes a magnetic
field of $B=15$~mT in the centre of the structure whose radius is 3.1~mm.
This field acts on the beam over the distance $l=6.7$~mm between irises and thus
causes a deflection on the beam of $Bl/B\rho=0.16$~mrad where $B\rho=0.6$~Tm is
the rigidity of the 180~MeV electron beam. We conclude that a current of a few
100~A between adjacent irises is consistent with the observed deflection.

\begin{acknowledgments}
The authors would like to acknowledge the support of the CTF3 operations team.
Especially we like to acknowledge the support of D. Gamba and A. Dubrovskiy for
data acquisition related issues. This work is supported by the 7th European
Framework program EuCard under Grant number 227579, the Knut and Alice
Wallenberg foundation and the Swedish Research Council.
\end{acknowledgments}

\bibliography{kick_prst}

\end{document}